\documentclass[runningheads]{llncs}
\usepackage{graphicx}
\usepackage[colorlinks=true,linkcolor=blue,urlcolor=blue,citecolor=blue]{hyperref}
\renewcommand\UrlFont{\color{blue}\rmfamily}

\newcommand{\partitle}[1]{\vspace{2mm}\noindent\textbf{#1}}

\newcommand{\modelname}{STACP\xspace}

\usepackage[ruled,vlined,linesnumbered]{algorithm2e}
\usepackage{subfig}
\usepackage{amsmath}
\usepackage{amssymb}
\usepackage{booktabs}
\usepackage{multirow}
\usepackage{multicol}
\usepackage{pgfplots}
\pgfplotsset{width=6cm,compat=1.14}
\usepackage{float}
\usepackage{rotating}
\usepackage{array,makecell}
\usepackage[export]{adjustbox}
\usepackage{arydshln}

\begin{document}
\title{Joint Geographical and Temporal\\  Modeling based on Matrix Factorization for Point-of-Interest Recommendation}
%
\titlerunning{Joint Geographical and Temporal Model}
%
\author{Hossein A. Rahmani\inst{1} \and
Mohammad Aliannejadi\inst{2}\thanks{Work done while Mohammad Aliannejadi was affiliated with Universit{\`a} della Svizzera italiana (USI).} \and
Mitra Baratchi\inst{3} \and \\
Fabio Crestani\inst{1}}
\authorrunning{H. A. Rahmani et al.}
\institute{Universit\`a della Svizzera Italiana, Lugano, Switzerland \\
\email{srahmani@znu.ac.ir, fabio.crestani@usi.ch}\\ \and
Univiersity of Amsterdam, Amsterdam, The Netherlands \\
\email{m.aliannejadi@uva.nl} \\ \and
Leiden University, Leiden, The Netherlands\\
\email{m.baratchi@liacs.leidenuniv.nl}}
\maketitle              
\begin{abstract}
With the popularity of Location-based Social Networks, Point-of-Interest (POI) recommendation has become an important task, which learns the users' preferences and mobility patterns to recommend POIs. Previous studies show that incorporating contextual information such as geographical and temporal influences is necessary to improve POI recommendation by addressing the data sparsity problem. However, existing methods model the geographical influence based on the physical distance between POIs and users, while ignoring the temporal characteristics of such geographical influences. In this paper, we perform a study on the user mobility patterns where we find out that users' check-ins happen around several centers depending on their current temporal state. Next, we propose a spatio-temporal activity-centers algorithm to model users' behavior more accurately. Finally, we demonstrate the effectiveness of our proposed contextual model by incorporating it into the matrix factorization model under two different settings: i) static and ii) temporal. To show the effectiveness of our proposed method, which we refer to as \modelname, we conduct experiments on two well-known real-world datasets acquired from Gowalla and Foursquare LBSNs. Experimental results show that the \modelname model achieves a statistically significant performance improvement, compared to the state-of-the-art techniques.
Also, we demonstrate the effectiveness of capturing geographical and temporal information for modeling users' activity centers and the importance of modeling them jointly. 

\keywords{Contextual Information  \and Point-of-Interest Recommendation \and Recommender System.}
\end{abstract}
\section{Introduction}
\label{sec:intro}
\vspace{-0.2cm}
With the availability of Location-based Social Networks (LBSNs) such as Yelp and Foursquare users can share their locations, experiences, and content associated with the Point-of-Interests (POIs) via check-ins. Employing the successes in the area of Recommender Systems (RSs), POI recommendation helps to improve the user experiences on LBSNs, suggesting POIs according to users' past check-in history. POI recommendation helps users explore new interesting POIs while helping businesses to increase their revenues by providing context-aware advertisements. As such, POI recommendation has attracted much attention from both research and industry \cite{liu2017experimental,aliannejadi2019joint,DBLP:conf/sigir/AliannejadiMC17}.

One of the most important challenges that limit the accuracy of POI recommendation is the data sparsity problem~\cite{adomavicius2005toward,stepan2016incorporating,rahmani2019category}. Numerous users are active on LBSNs with millions of POIs already being listed on these platforms. However, in practice, users are able to only visit a very limited number of POIs. Hence, the user-POI matrix used in different Collaborative Filtering (CF) approaches becomes sparse, limiting the attainable recommendation accuracy \cite{aliannejadi2018personalized,rahmani2019lglmf}. To address this problem, several studies have incorporated contextual information such as geographical and temporal influences separately into their model \cite{baral2016geotecs,li2015rank,yuan2013time,zhang2013igslr}. For example, relevant studies have tried to incorporate geographical \cite{ye2011exploiting,cheng2012mgm,li2015rank} and temporal influences \cite{griesner2015poi,yao2016poi,li2017time} in their proposed model. Moreover, as argued in \cite{cheng2012mgm,cheng2016unified,cho2011friendship}, users commonly check in to POIs around several geographical centers. While modeling these centers it is assumed that they are static and do not change according to temporal information. This assumption may not be correct and taking into account both geographical and temporal influences might help to model the users' behavior with a higher accuracy. 
For instance, if we consider working time and leisure time, this suggests that users tend to explore POIs around their activity centers in leisure time, while they prefer to visit the same locations more often while they are at work. For example, a person would go to the same restaurant every day to have lunch while working during the weekdays. However, the same user might decide to visit a more diverse set of POIs and visit new places while on holidays (i.e., leisure time). Therefore, the users' check-in behavior and activity centers are dependent on their temporal states (e.g., working time vs. leisure time).

To elaborate more, in Figure~\ref{fig:intown}, we have depicted a randomly selected user's check-ins from the Gowalla dataset \cite{liu2017experimental} during working and leisure time.
As seen, this user follows a temporal center-based check-in pattern; that is, the activity centers are different at different temporal states. Also, while we compare Figure~\ref{fig:all} with Figures~\ref{fig:working} and \ref{fig:leisure}, we see that the activity centers are different from each temporal state, compared to all the check-ins. 
Based on these observations, we conclude that joint modeling of geographical and temporal information is an effective approach for defining users' activity centers. In this paper, we take a step for joint modeling of geographical and temporal information. Our contributions can be summarized as follows:

\begin{figure}[!tbp]
  \centering
  \subfloat[All check-ins]{\includegraphics[width=0.3\textwidth]{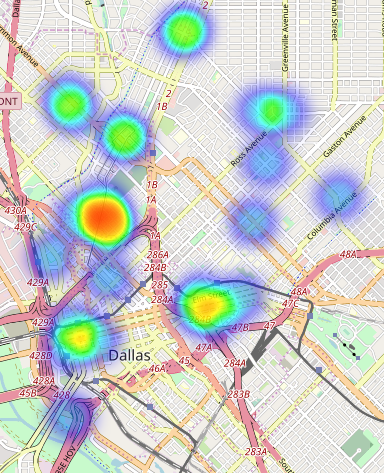}\label{fig:all}}
  \hfill
  \subfloat[Working check-ins]{\includegraphics[width=0.3\textwidth]{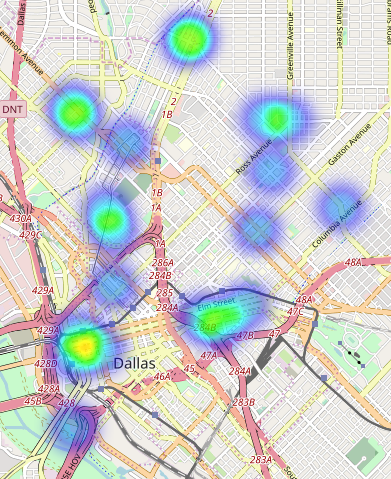}\label{fig:working}}
  \hfill
  \subfloat[Leisure check-ins]{\includegraphics[width=0.3\textwidth]{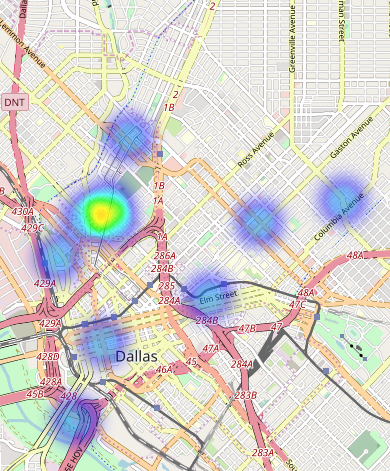}\label{fig:leisure}}
  \caption{A typical user's spatio-temporal activity centers from the Gowalla dataset. As we see, Figure \ref{fig:all} shows all check-ins of the user whereas Figures \ref{fig:working} and \ref{fig:leisure} show the check-ins for working and leisure time are focused in different centers. (best viewed in color)}
  \label{fig:intown}
\end{figure}

\begin{itemize}
    \item We propose a novel contextual model that jointly considers both geographical and temporal information.
    \item We propose a spatio-temporal activity-centers model that consider users' center-based behavior in different temporal states.
    \item We propose static and temporal MF models to study the users' preference and behavior both in static and temporal manners. In the static MF, we train the model on the whole user-POI matrix, whereas in the temporal MF, we train the model using different user-POI matrices for every time slot.
    \item We address the data sparsity problem by incorporating the proposed contextual model into the traditional MF model and propose a novel MF framework.
\end{itemize}

We conduct several experiments on two well-known real-world datasets, namely, Gowalla and Foursquare, demonstrating the improvement of the proposed method in the accuracy of POI recommendation compared to a number state-of-the-art approaches. Our experiments show that joint modeling of the geographical and temporal influence improves the performance of POI recommendation. Finally, to enable reproducibility of the results, we have made our codes open source.\footnote{\UrlFont{https://github.com/rahmanidashti/STACP}}

\section{Related Work}
\label{sec:relatedwork}
\vspace{-0.2cm}
POI recommendation plays an essential role in improving LBSNs user experience. Much work has been carried out in this area based on the core idea behind recommendation systems, assuming that \textit{users with similar behavioral histories tend to act similarly}~\cite{ye2011exploiting}. Collaborative Filtering (CF-based) recommendation approaches aim to base recommendations on the similarity between users and items~\cite{ference2013location,griesner2015poi}. POI recommendation considers a large number of available POIs (items) while a single user can only visit a few of them. Hence, CF-based approaches applied to POI recommendation often suffer from the data sparsity problem. This leads to poor performance in POI recommendation. Many studies have tried to address the data sparsity problem of CF approaches incorporating additional information into the model~\cite{ye2011exploiting,li2015rank,liu2017experimental,aliannejadi2017personalized}. As the users' check-in behavior follows a spatio-temporal mobility pattern, much work has incorporated this critical information. Considering users' movement trajectories between POIs, many of the previous studies have shown that geographical influence is one of the most important factors in POI recommendation, ~\cite{rahmani2019lglmf,liu2017experimental}. 

More specifically, Ye et al.~\cite{ye2011exploiting} argued that a user's check-in behavior is affected by the geographical influence of POIs, following the power-law distribution and proposed a unified POI recommender system incorporating spatial and social influences to address the data sparsity problem. Ference et al.~\cite{ference2013location} took into consideration several factors such as user preference, geographical proximity, and social influences for out-of-town POI recommendation. This work, however, did not take into account users' temporal information and in-town users' behavior. Cheng et al.~\cite{cheng2012mgm,cheng2016unified} modeled users' check-ins via center-based Gaussian distribution to capture users' movement patterns based on the assumption that users' movements consist of several centers. Li et al.~\cite{li2015rank}, in another work, modeled the POI recommendation task as a pairwise ranking loss, where they exploited the geographical information using an extra factor matrix. Zhang et al.~\cite{zhang2013igslr} proposed a method that considered the geographical influence on each user separately. To this end, they proposed a model based on kernel density estimation of the distance distributions between POI check-ins per user. Aliannejadi et al.~\cite{aliannejadi2017venue} proposed a ranking model and predicted the appropriateness of a POI given a user's context into the ranking process. Yuan et al.~\cite{yuan2016joint} addressed the data sparsity problem based on the idea that users tend to rank higher those POIs that are geographically closer to their visited POIs. Guo et al.~\cite{guo2019location} proposed a location neighborhood-aware weighted matrix factorization model to exploit the location perspective that incorporates the geographical relationships among POIs. More recently, Aliannejadi et al.~\cite{aliannejadi2019joint} proposed a two-phase collaborative ranking algorithm for POI recommendation that takes into account the geographical influence of POIs in the same neighborhood.

Another line of research studies the temporal influence on users' preferences \cite{ding2005time,yuan2013time,zhao2017geo}. Temporal information has been shown to improve POI recommendation accuracy and alleviate the problem of data sparsity \cite{liu2017experimental}. Griesner et al.~\cite{griesner2015poi} proposed an approach to integrate temporal influences into matrix factorization. Gao et al.~\cite{gao2013exploring} computed the similarity between users by dividing users' check-ins into different hourly time slots and finding the same POIs at the same time slots in their check-in history to train a user-based CF model. Yao et al.~\cite{yao2016poi} matched the temporal regularity of users with the popularity of POIs to improve a factorization-based algorithm. Le et al.~\cite{li2017time} proposed a time-aware personalized model adopting a fourth-order tensor factorization-based ranking, which enables to capture short-term and long-term preferences. Yuan et al.~\cite{yuan2013time} preserved the similarity of personal preference in consecutive time slots by considering different latent variables at each time slot per user. Zhao et al.~\cite{zhao2016stellar} proposed a latent ranking method to model the temporal interactions among users and POIs explicitly. In particular, the proposed model builds upon a ranking-based pairwise tensor factorization framework.

These previous approaches mainly explored the geographical and temporal information separately. Differently from these studies, our work addresses the data sparsity problem by jointly modeling the geographical and temporal contextual information. Moreover, the previous research modeled users' center-based behavior based on geographical influence. In contrast, we consider the formation of spatio-temporal activity centers for each user. Therefore, we model the users' center-based behavior based on different temporal states. 

\section{Proposed Approach}
In this section, we propose a \textbf{S}patio-\textbf{T}emporal \textbf{A}ctivity \textbf{C}enter \textbf{P}OI recommendation model called \modelname, which models users' preference and users' context together. In the users' preference model, we design two preference functions for each user to consider both static and temporal users' preferences in the model. Moreover, in the users' context model, we incorporate the influence of geographical and temporal information jointly. In what follows, we first describe an overview of our \modelname model and further explain how each part is implemented and which challenges are addressed at each part.

Formally, let $\mathcal{U}=\{u_1,u_2,u_3,...,u_m\}$ be the set of users and $\mathcal{L}=\{l_1,l_2,l_3,...,l_n\}$ be the set of POIs. Further, let $m$ and $n$ be the number of users and POIs, respectively. Then, the users visit-frequency can be encoded in $R_{m\times{n}}$, where entries $r_{u,l} \in R$ can represent the previous POI check-ins of user $u \in \mathcal{U}$ to POI $l \in \mathcal{L}$. Also, $\mathcal{L}_u$ shows all POIs checked-in by user $u$. To address the data sparsity problem and explore the contextual influence we need to fuse the users' context with the users' preference model in a fusion framework. We fuse users' static and temporal preferences on a POI and the score of whether a user will visit that place based on our contextual influence model. \modelname is proposed to estimate the recommendation score that a user $u$ visits a POI $l$ as follows:

\begin{equation}
    \modelname_{u,l}={U_u^TL_l}\times{P(u,l|C_{u,t})}\times{\hat{R_{u,l}}}
    \label{eq:fusion}
\end{equation}

\noindent
where ${U_u^TL_l}$ and $\hat{R_{u,l}}$ respectively denote static and temporal users' preference model, and $P(u,l|C_{u,t})$ denotes the users' context model.

In the following, we first introduce the user context model, where we show users' behavior in a joint model of geographical and temporal influences. Moreover, we propose our temporal center allocation method. Finally, we describe our users' static and temporal preference model.

\partitle{Spatio-Temporal Activity Centers.}
\label{sec:tempcentermodel}
As shown in Figure~\ref{fig:intown}, users' behaviors are center-based and these centers are different based on the periodicity of temporal information (see Figure \ref{fig:working} and \ref{fig:leisure}). This phenomenon points to the shortcoming of the previous geographical and temporal models that considered geographical and temporal influences separately. As shown previously, the second characteristic of users' behavior is that users tend to visit POIs that are near their current centers. We apply these two characteristics jointly to model users' check-in behavior and propose the spatio-temporal activity-centers model. That is, the score of a user $u$, visiting a POI $l$, given the temporal multi-center set $C_{u,t}$ of user $u$ in time $t$ and temporal state $T$, is defined as follows:
\begin{equation}
    P(u,l|C_{u,t};T)=\sum_{c_{u,t}}^{|C_{u,t}|}{\frac{1}{dist(l,c_{u,t})}\frac{freq_{c_{u,t}}}{\sum_{i\in{C_{u,t}}}{freq_{i}}}}
    \label{eq:centerscore}
\end{equation}
where $l$ denotes a POI and $C_{u,t}$ is the set of centers for the user $u$ in time $t$, given the temporal state $T$. For each center, calculating \eqref{eq:centerscore} consists of the multiplication of two terms. The first term determines the score of the POI $l$ belonging to the center $c_{u,t}$, which is related to the distance between the POI $l$ and the center $c_{u,t}$. The second term denotes the effect of check-in frequency $freq_{c_{u,t}}$, on the center $c_{u,t}$. 

Further, we define the multi-center activity function $P(u,l|C_{u,t})$ as a linear interpolation under two temporal states, as follows:
\begin{equation}
    P(u,l|C_{u,t}) = \lambda \times P(u,l|C_{u,t}; WT) + (1-\lambda) \times P(u,l|C_{u,t}; LT)
\end{equation}
\noindent
where we consider it for working time $P(u,l|C_{u,t}; WT)$ and leisure time $P(u,l|C_{u,t}; LT)$ where $\lambda$ shows the impact of each temporal state. The model can be generalized to define other temporal states. For example, we could apply it for weekday/weekend, monthly, or daily patterns.

\partitle{Activity Center Allocation.}
\label{sec:tempcenterfinder}
As argued earlier, the users' activities follow a center-based pattern. Furthermore, these centers are different depending on the temporal state. To model the users' behavior in a spatio-temporal manner, we propose a temporal multi-center clustering algorithm among each user's check-ins based on the Pareto principle \cite{hafner2001pareto}, as the most visited POIs account for a few users. First, for each user $u$ and temporal state $t$, we rank all POIs $L_u$ according to the check-in frequency. Next, we select the most visited POI and combine all other visited POIs within $d$ kilometers from the selected POI, to create a region. Let $N_u$ be the user $u$'s total check-in numbers, $r$ be the current region and $N_{r,u}$ be the total check-in number of current region of user $u$. To decide if a center should be added to the user's profile, we consider a threshold of $\alpha$. A new center is considered if $\frac{N_{r,u}}{N_u} > \alpha$.
We repeat this procedure until we cover all of the user $u$'s checked-in POIs.

\partitle{Static and Temporal Users' Preferences.}
\label{sec:prefrence}
To model the user's preference based on check-in data, we apply Matrix Factorization (MF) in two ways: a static model of user's preference (SMP) and a temporal model of user's preference (TMP). In SMP, we consider the traditional matrix factorization method to model the static behavior of users. The goal of MF is to find two low-rank matrices $U \in \mathbb{R}^{K\times|\mathcal{U}|}$ and $L \in \mathbb{R}^{K\times|\mathcal{L}|}$ based on the frequency matrix $R$ such that $R \approx U^{T}L$. The predicted recommendation score of a user $u$, like a POI $l$, is determined by:
\begin{equation}
    P_{u,l} \propto U_u^TL_l
\end{equation}
\noindent
via solving the following optimization problem which places Beta distributions as priors on the latent matrices $U$ and $L$, while defining a Poisson distribution on the frequency:
\begin{equation}
\begin{split}
    min_{\{U,L|R\}} = \sum_{i=1}^{|\mathcal{U}|}\sum_{k=1}^{K}((\sigma_k-1)\ln(U_{ik}/\rho_k)-U_{ik}/\rho_k) \\ +
    \sum_{j=1}^{|\mathcal{L}|}\sum_{k=1}^{K}((\sigma_k-1)\ln(L_{jk}/\rho_k)-L_{jk}/\rho_k) \\ +
    \sum_{i=1}^{|\mathcal{U}|}\sum_{j=1}^{|\mathcal{L}|}((R_{ij}\ln(U^TL)_{ij}-(U^TL)_{ij})+c
\end{split}
\end{equation}
\noindent
where $\sigma=\{\sigma_1,...,\sigma_K\}$ and $ \rho=\{\rho_1,...,\rho_K\}$ are parameters for Beta distributions, and $c$ is a constant term. 
In TMP, to model the temporal behavior of users, inspired by \cite{gao2013exploring}, we divide the original user-POI frequency matrix $R$ into $t$ sub-matrices according to the different temporal states $T$. Then each sub-matrix only containing check-in actions that happened at the corresponding temporal state. For example, we can consider $t=2$ for working time and leisure time in our case. Then, we apply MF on each $R_t$ to compute user $u$'s preference on POI $l$ at time $t$. Finally, we sum them into $\hat{R}$, representing the user check-in preferences of each POI. It should be mentioned that a more advanced method of automated periodic pattern extraction from spatio-temporal data as proposed in \cite{periodcpatterns} can also be used for a more data-informed decision to be made for the parameter ($t$).

\vspace{-0.3cm}
\section{Experiments}
\vspace{-0.2cm}
In this section, several experiments are conducted to compare the performance of \modelname with the other state-of-the-art POI recommendation methods.

\subsection{Experimental Setup}
\partitle{Datasets.}
We use two real-world check-in datasets from Gowalla and Foursquare provided by \cite{liu2017experimental}\footnote{\UrlFont{http://spatialkeyword.sce.ntu.edu.sg/eval-vldb17/}}. The Gowalla dataset consists of 620,683 number of world-wide check-ins made by 5,628 number of users on 31,803 POIs with 99.78\% sparsity in the period of Feb. 2009 to Oct. 2010. The Foursquare dataset, on the other hand, includes 512,523 check-ins made by 7,642 users on 28,483 POIs with 99.87\% sparsity in the United States from Apr. 2012 to Sep. 2013. Every check-in contains a user-id, POI-id, time, and geographical coordinates.

\partitle{Evaluation Metrics.} To evaluate the performance of the recommendation methods, we used three evaluation metrics: Precision@N, Recall@N, and nDCG@N with $N \in \{10, 20\}$. We partition each dataset into training data, validation data, and test data. For each user, we use the earliest 70\% check-ins as training data, the most recent 20\% check-ins as test data, and the remaining 10\% as validation data. We determine the statistically significant differences in the results using the two-tailed paired t-test at a 95\% confidence interval ($p < 0.05$).

\begin{table*}[t]
\settowidth\rotheadsize{FouesquareDataset}
  \caption{Performance comparison with baselines in terms of Precision@$N$, Recall@$N$, and nDCG@$N$ for $N \in \{10,20\}$ on Gowalla and Foursquare. The superscripts $\dagger$ and $\ddagger$ denote significant improvements compared to baselines and model variations, respectively ($p < 0.05$).}
  \label{tab:results}
  \begin{adjustbox}{max width=\textwidth}
      \begin{tabular}{ll@{\quad}r@{\quad}rr@{\quad}@{\quad}r@{\quad}rr@{\quad}@{\quad}r@{\quad}rr}
        \toprule
        \multirow{2}{*}{} & \multirow{2}{*}{} & \multicolumn{2}{c}{\textbf{Precision}} && \multicolumn{2}{c}{\textbf{Recall}} && \multicolumn{2}{c}{\textbf{nDCG}} \\
        \cmidrule{3-4} \cmidrule{6-7} \cmidrule{9-10}
        & & \multicolumn{1}{c}{\textbf{@10}} & \multicolumn{1}{c}{\textbf{@20}} && \multicolumn{1}{c}{\textbf{@10}} & \multicolumn{1}{c}{\textbf{@20}} && \multicolumn{1}{c}{\textbf{@10}} & \multicolumn{1}{c}{\textbf{@20}} \\
        \midrule
        \multirow{8}{*}[0ex]{\rothead{\textbf{Gowalla}}} & TopPopular & 0.0192 & 0.0146 && 0.0176 & 0.0270 && 0.0088 & 0.0079 \\
        & PFM & 0.0181 & 0.0143 && 0.0161 & 0.0252 && 0.0077 & 0.0068 \\
        & PFMMGM & 0.0240 & 0.0207 && 0.0258 & 0.0442 && 0.0140 & 0.0144 \\
        & LRT & 0.0249 & 0.0182 && 0.0220 & 0.0321 && 0.0105 & 0.0093 \\
        & PFMPD & 0.0217 & 0.0184 && 0.0223 & 0.0373 && 0.0099 & 0.0101 \\
        & LMFT & 0.0315 & 0.0269 && 0.0303 & 0.0515 && 0.0157 & 0.0150 \\
        & iGLSR & 0.0297 & 0.0242 && 0.0283 & 0.0441 && 0.0153 & 0.0145 \\
        & MLP & 0.0243 & 0.0215 && 0.0237 & 0.0396 && 0.0982 & 0.0127 \\
        & Rank-GeoFM & 0.0352 & 0.0297 && 0.0379 & 0.0602 && 0.0187 & 0.0179 \\
        & L-WMF & 0.0341 & 0.0296 && 0.0351 & 0.0582 && 0.0183 & 0.0178 \\
        \hdashline
        & \modelname-NoCTX & 0.0219 & 0.0167 && 0.1920 & 0.0293 && 0.0092 & 0.0081 \\
        & \modelname-NoTC & 0.0282 & 0.0236 && 0.0281 & 0.0457 && 0.0147 & 0.0151 \\
        & \textbf{\modelname} & \textbf{0.0383}$^{\dagger\ddagger}$ & \textbf{0.0318}$^{\ddagger}$ && \textbf{0.0404}$^{\dagger\ddagger}$ & \textbf{0.0651}$^{\dagger\ddagger}$ && \textbf{0.0212}$^{\dagger\ddagger}$ & \textbf{0.0211}$^{\dagger\ddagger}$ \\
      \midrule
        \multirow{8}{*}[0ex]{\rothead{\textbf{Foursquare}}} & TopPopular & 0.0200 & 0.0155 && 0.0272 & 0.0429 && 0.0114 & 0.0121 \\
        & PFM & 0.0213 & 0.0154 && 0.0290 & 0.0424 && 0.0125 & 0.0129 \\
        & PFMMGM & 0.0170 & 0.0150 && 0.0283 & 0.0505 && 0.0109 & 0.0126 \\
        & LRT & 0.0199 & 0.0155 && 0.0265 & 0.0425 && 0.0117 & 0.0124 \\
        & PFMPD & 0.0214 & 0.0155 && 0.0290 & 0.0426 && 0.0124 & 0.0128 \\
        & LMFT & 0.0241 & 0.0194 && 0.0359 & 0.0568 && 0.0150 & 0.0161 \\
        & MLP & 0.0248 & 0.0204 && 0.0309 & 0.0373 && 0.0135 & 0.0152 \\
        & Rank-GeoFM & 0.0263 & 0.0241 && 0.0399 & 0.0625 && 0.0183 & 0.0197 \\
        & L-WMF & 0.0248 & 0.0197 && 0.0387 & 0.0591 && 0.0162 & 0.0174 \\
        \hdashline
        & \modelname-NoCTX & 0.0207 & 0.0184 && 0.0196 & 0.0285 && 0.0094 & 0.0085 \\
        & \modelname-NoTC & 0.0242 & 0.0217 && 0.0331 & 0.0552 && 0.0142 & 0.0157 \\
        & \textbf{\modelname} & \textbf{0.0312}$^{\dagger\ddagger}$ & \textbf{0.0285}$^{\dagger\ddagger}$ && \textbf{0.0453}$^{\dagger\ddagger}$ & \textbf{0.0671}$^{\dagger\ddagger}$ && \textbf{0.0203}$^{\ddagger}$ & \textbf{0.0227}$^{\dagger\ddagger}$ \\
      \bottomrule
    \end{tabular}
\end{adjustbox}
\end{table*}

\partitle{Comparison Methods.}
We compared the proposed \modelname model with the state-of-the-art POI recommendation approaches that consider geographical or temporal influences in the recommendation process. The details of the compared methods are listed below:
\begin{itemize}
    \item \textbf{TopPopular} \cite{dacrema2019we}: A simple and non-personalized method that recommends the most popular POIs to users. Popularity is measured by the number of check-ins.
        \item \textbf{PFM} \cite{ma2011probabilistic}: A MF method, which can model the frequency data directly. PFM places Beta distributions as priors on the latent matrices $U$ and $V$, while defining a Poisson distribution on the frequency.
    \item \textbf{PFMMGM} \cite{cheng2012mgm}: A method based on the observation that a user's check-ins follow a Gaussian distribution that combines geographical and social influence with MF.
    \item \textbf{LRT} \cite{gao2013exploring}: A method that incorporates temporal information in a latent ranking model and learns the user's preferences based on temporal influence.
    \item \textbf{PFMPD}: A geographical method using the Power-law distribution \cite{ye2011exploiting} that models people's tendency to visit nearby POIs. We integrate this geographical model with the Probabilistic Factor Model (PFM).
    \item \textbf{LMFT} \cite{stepan2016incorporating}: A method that applies temporal information on the user's recent activities and multiple visits to a POI.
    \item \textbf{iGLSR}\footnote{We evaluate iGLSR only on Gowalla as we do not have access to the social data of the Foursquare dataset.} \cite{zhang2013igslr}: A method that personalizes social and geographical influences on POI recommendation using a Kernel Density Estimation (KDE) approach.
    \item \textbf{Rank-GeoFM} \cite{li2015rank}: A ranking-based MF model that includes the geographical influence of neighboring POIs while learning user preference rankings for POIs.
    \item \textbf{MLP} \cite{he2017neural}: A component of the NeuMF framework that models the user-POI interaction using the concatenation of latent factors via Multi-layers Neural Network.
    \item \textbf{L-WMF} \cite{guo2019location}: A location neighborhood-aware weighted probabilistic matrix factorization model. L-WMF incorporates the geographical relationships among POIs as regularization to exploit the geographical characteristics from a location perspective.
    \item \textbf{STMCP-NoCTX}: A variation of our model which excludes the contextual model. We include this model as a baseline to demonstrate the effectiveness of our contextual model. 
    \item \textbf{STMCP-NoTC}: A variation of our model in which we remove the temporal states and consider geographical centers without temporal differences. We include this model to show the effectiveness of temporal centers in our model.
\end{itemize}

\begin{figure}[t]
  \centering
  \subfloat[Gowalla]{
  \includegraphics[scale=0.35]{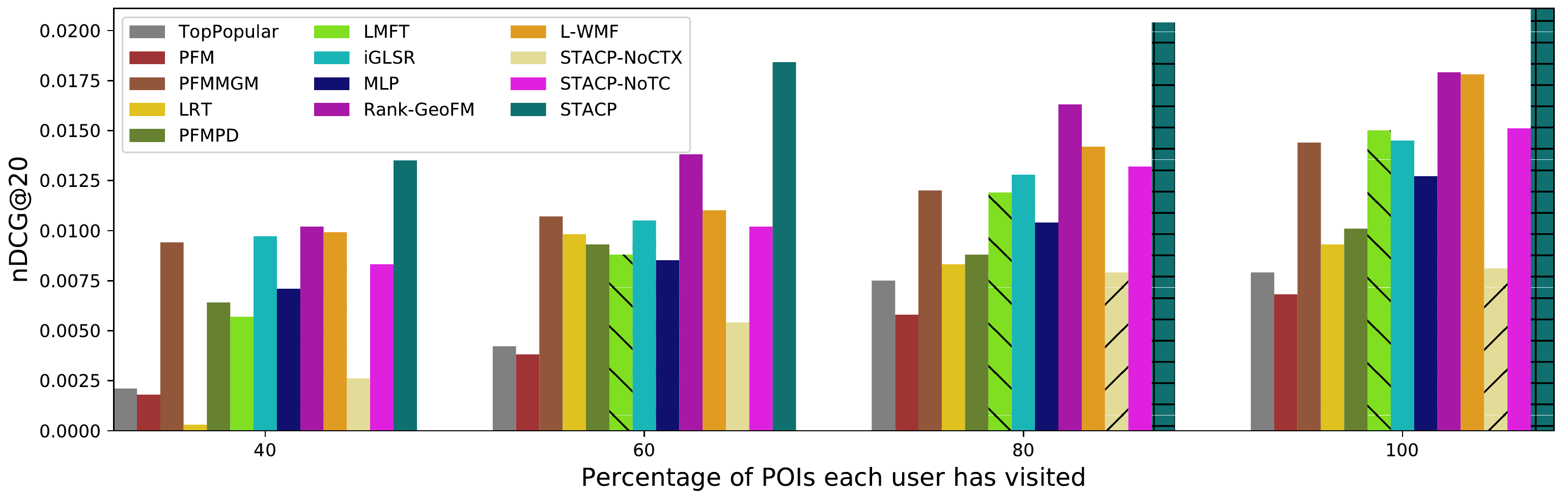}
  \label{fig:sparsityGowalla}
  }
  \hfill
  \subfloat[Foursquare]{
  \includegraphics[scale=0.35]{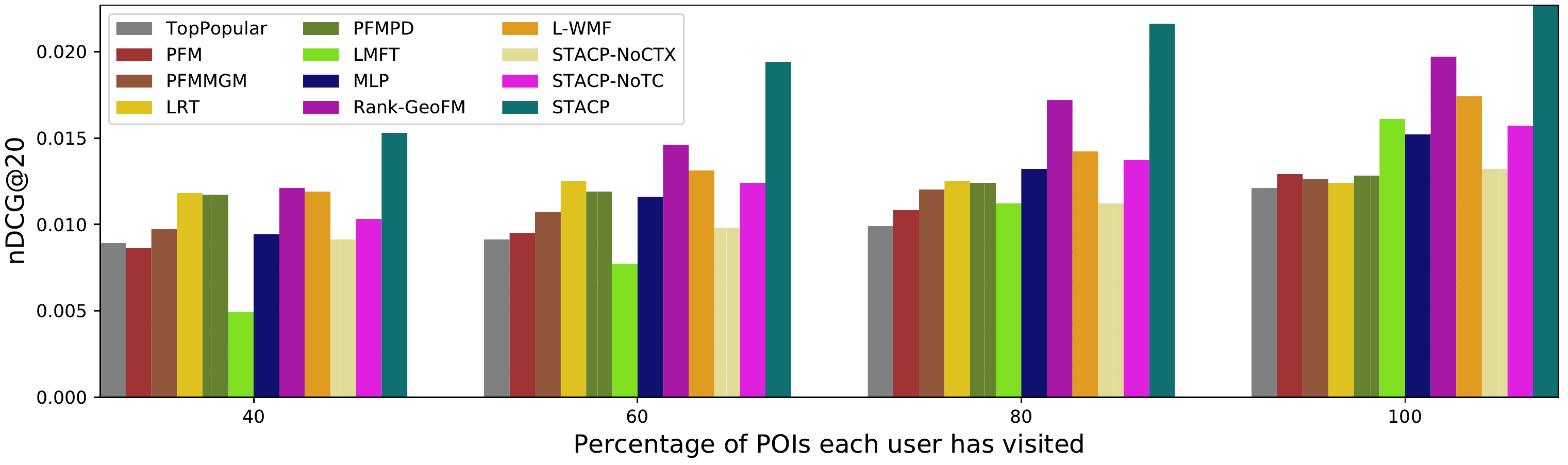}
  \label{fig:sparsityFoursqaure}
  }
  \caption{Effect on nDCG@20 by varying the percentage of POIs that each user has visited for (a) Gowalla and (b) Foursquare.}
  \label{fig:sparsity}
\end{figure}
\subsection{Results}
\partitle{Performance evaluation against compared methods.}
Table \ref{tab:results} shows the results of experiments based on the Gowalla and Foursquare datasets. As seen, \modelname obtains the best performance compared to the other POI recommendation methods in terms of all evaluation metrics on both of the datasets. Dacrema et al. in~\cite{dacrema2019we} show that even some state-of-the-art deep-learning-based methods are not able to outperform a non-personalized method such as TopPopular. This is the reason why we have also selected this method as a baseline to compare with \modelname. It seems that in comparison to two non-personalized baseline methods, TopPopular and PFM, our method achieves significantly better performance. Comparing with other geographical-based methods, PFMPD, PFMMGM, iGLSR, and L-WMF, it is seen that \modelname followed by Rank-GeoFM perform best. One reason for the performance of Rank-GeoFM is that it considers the geographical neighborhoods as a major element in the factorization method. Also, Rank-GeoFM takes a ranking approach to modeling the interactions between users and POIs. This means that instead of considering a point-wise loss function, it applies a pairwise loss function in geographical factorization.

Results show that \modelname beats all geographical-based methods in terms of all metrics for all different values of $N$ in both datasets. This is expected as the previous models only consider the basic idea of the geographical influence that users tend to visit nearby POIs. Also, our proposed model outperforms the LRT and LMFT that modeled temporal information to improve the accuracy of POI recommendation. The reason is that these methods do not consider the geographical information. Compared to the neural baseline, MLP, the improvements of \modelname in terms of Recall@20 and nDCG@20 on Gowalla dataset are $65\%$ and $66\%$, respectively. This shows the effectiveness of our users' preference and users' context models, which considers both geographical and temporal information jointly to model users' activity centers.

\partitle{Effect of Activity Centers.}
In this experiment, we compare the performance of \modelname with its variation where we only consider geographical information in allocating activity centers (i.e., \modelname-NoTC). Therefore, we remove temporal states $t$ from equation \eqref{eq:centerscore}. The goal is to demonstrate the effect of the spatio-temporal activity centers on the performance of \modelname. As seen in Table \ref{tab:results}, \modelname exhibits a significant improvement over \modelname-NoTC in terms of all evaluation metrics for both datasets. We see that \modelname improves \modelname-NoTC by 42\% in terms of Recall@20. This indicates that users follow a spatio-temporally centered behavior. This objectively validates our analysis in Section \ref{sec:intro}, in which users' centers are different based on the different temporal states.

\partitle{Effect of Contextual Model.}
Next, we study the effect of the contextual model. To this end, we compare the performance of \modelname with its variation where no contextual information is used while training the model (i.e., \modelname-NoCTX). In other words, in this experiment equation \eqref{eq:centerscore} is excluded from equation \eqref{eq:fusion}. As seen in Table \ref{tab:results}, a statistically significant improvement of \modelname over \modelname-NoCTX is observed in terms of all evaluation metrics on both datasets. This observation suggests that using contextual information enables \modelname to model the users' behavior more accurately. Moreover, it indicates that by incorporating the contextual information, we can address the data sparsity problem. 

\partitle{Effect of number of visited POIs.}
In this experiment, our goal is to study the effect of data size on the performance of our model. As such, we train \modelname, as well as all the baseline methods with different data sizes. To do so, for each user, we only consider a certain percentage of visited POIs in the training set randomly, ranging from 40\% to 100\%.
We see in Figures~\ref{fig:sparsityGowalla} and \ref{fig:sparsityFoursqaure} the performance of \modelname and all baseline models in terms of nDCG@$20$ for different training data sizes. The results show that \modelname is more effective in comparison with the baselines as the size of the training data varies, indicating that it addresses the data sparsity problem more effectively. As we see in Figure \ref{fig:sparsityGowalla}, when we change the data size from 100\% to 40\% on Gowalla, the performance of \modelname decreases by about 35\%, while for the competitor baseline method Rank-GeoFM the value of decrease is 45\%. This shows that \modelname is more robust when we do not have access to enough data from users. More interesting, the performance of LMFT, a temporal-information-based competitor baseline, decreases by 65\% when the size of data changes to 40\%. This indicates the unsuitability of the methods that only consider the temporal information. Also, it is worth noting that we observe a more robust behavior of \modelname compared to the \modelname-NoCN. Thus, the proposed context model enables \modelname to deal with noise and data sparsity more effectively. This is clearer when \modelname outperforms the best competitor baseline (i.e., Rank-GeoFM) with a larger margin, 33\%, in terms of nDCG@$20$ on Gowalla.

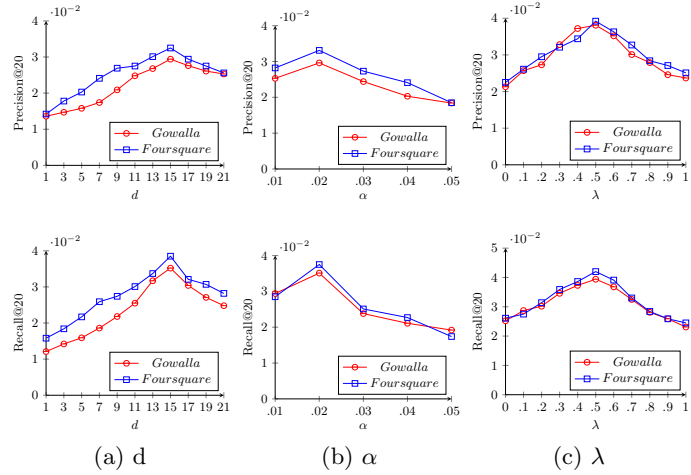
\begin{figure}[t]
	\centering
	\captionsetup[subfigure]{labelformat=empty}
	\subfloat{%
		\label{fig:d-pre}%
		\resizebox{.25\textwidth}{!}{
		\begin{tikzpicture}
        \begin{axis}[
            axis lines = left,
            xlabel = $d$,
            ylabel = {Precision@$20$},
            legend pos=south east,
            xmin=1, xmax=21,
            ymin=0, ymax=0.04,
            xtick={1,3,5,7,9,11,13,15,17,19,21},
            ytick={0,0.01,0.02,0.03,0.04,0.05,0.06}
        ]
        \addplot [
            color=red,
            mark=halfcircle
        ]
        coordinates{
        (1,0.0136)(3,0.0147)(5,0.0158)(7,0.0174)(9,0.0209)(11,0.0248)(13,0.0268)(15,0.0294)(17,0.0276)(19,0.0261)(21,0.0253)
        };
        \addlegendentry{$Gowalla$}
        \addplot [
            color=blue,
            mark=square
            ]
            coordinates{
            (1,0.0142)(3,0.0178)(5,0.0203)(7,0.0241)(9,0.0269)(11,0.0275)(13,0.0301)(15,0.0325)(17,0.0294)(19,0.0275)(21,0.0256)
            };
        \addlegendentry{$Foursquare$}
        \end{axis}
        \end{tikzpicture}}}
    \subfloat{%
		\label{fig:a-pre}%
		\resizebox{.25\textwidth}{!}{
		\begin{tikzpicture}
        \begin{axis}[
            axis lines = left,
            scaled x ticks = false,
            xlabel = $\alpha$,
            ylabel = {Precision@$20$},
            legend pos=south east,
            xmin=0.01, xmax=0.05,
            ymin=0, ymax=0.04,
            xtick={0.01,0.02,0.03,0.04,0.05},
            xticklabels={.01,.02,.03,.04,.05},
            ytick={0,0.01,0.02,0.03,0.04}
        ]
        \addplot [
            color=red,
            mark=halfcircle
        ]
        coordinates{
        (0.01,0.0253)(0.02,0.0296)(0.03,0.0244)(0.04,0.0203)(0.05,0.0184)
        };
        \addlegendentry{$Gowalla$}
        \addplot [
            color=blue,
            mark=square
            ]
            coordinates{
            (0.01,0.0282)(0.02,0.0331)(0.03,0.0273)(0.04,0.0241)(0.05,0.0185)
            };
        \addlegendentry{$Foursquare$}
        \end{axis}
        \end{tikzpicture}}}
    \subfloat{%
		\label{fig:g-pre}%
		\resizebox{.25\textwidth}{!}{
		\begin{tikzpicture}
        \begin{axis}[
            axis lines = left,
            xlabel = $\lambda$,
            ylabel = {Precision@$20$},
            legend pos=south east,
            xmin=0, xmax=1,
            ymin=0, ymax=0.04,
            xtick={0,.1,.2,.3,.4,.5,.6,.7,.8,.9,1},
            xticklabels={0,.1,.2,.3,.4,.5,.6,.7,.8,.9,1},
            ytick={0,0.01,0.02,0.03,0.04}
        ]
        \addplot [
            color=red,
            mark=halfcircle
        ]
        coordinates{
        (0,0.0213)(0.1,0.0257)(0.2,0.0273)(0.3,0.0328)(0.4,0.0372)(0.5,0.0381)(0.6,0.0352)(0.7,0.0301)(0.8,0.0279)(0.9,0.0246)(1,0.0237)
        };
        \addlegendentry{$Gowalla$}
        \addplot [
            color=blue,
            mark=square
            ]
            coordinates{
            (0,0.0225)(0.1,0.0261)(0.2,0.0295)(0.3,0.0321)(0.4,0.0344)(0.5,0.0392)(0.6,0.0363)(0.7,0.0327)(0.8,0.0284)(0.9,0.0271)(1,0.0251)
            };
        \addlegendentry{$Foursquare$}
        \end{axis}
        \end{tikzpicture}}}
    
    \subfloat[][(a) d]{%
		\label{fig:d-rec}%
		\resizebox{.25\textwidth}{!}{
		\begin{tikzpicture}
        \begin{axis}[
            axis lines = left,
            xlabel = $d$,
            ylabel = {Recall@$20$},
            legend pos=south east,
            xmin=1, xmax=21,
            ymin=0, ymax=0.04,
            xtick={1,3,5,7,9,11,13,15,17,19,21},
            ytick={0,0.01,0.02,0.03,0.04}
        ]
        \addplot [
            color=red,
            mark=halfcircle
        ]
        coordinates{
        (1,0.0121)(3,0.0142)(5,0.0159)(7,0.0186)(9,0.0218)(11,0.0255)(13,0.0317)(15,0.0352)(17,0.0304)(19,0.0271)(21,0.0248)
        };
        \addlegendentry{$Gowalla$}
        \addplot [
            color=blue,
            mark=square
            ]
            coordinates{
            (1,0.0158)(3,0.0184)(5,0.0217)(7,0.0259)(9,0.0274)(11,0.0301)(13,0.0337)(15,0.0385)(17,0.0321)(19,0.0307)(21,0.0282)
            };
        \addlegendentry{$Foursquare$}
        \end{axis}
        \end{tikzpicture}}}
	\subfloat[][(b) $\alpha$]{%
		\label{fig:a-rec}%
		\resizebox{.25\textwidth}{!}{
		\begin{tikzpicture}
        \begin{axis}[
            axis lines = left,
            scaled x ticks = false,
            xlabel = $\alpha$,
            ylabel = {Recall@$20$},
            legend pos=south east,
            xmin=0.01, xmax=0.05,
            ymin=0, ymax=0.04,
            xtick={0.01,0.02,0.03,0.04, 0.05},
            xticklabels={.01,.02,.03,.04,.05},
            ytick={0,0.01,0.02,0.03,0.04}
        ]
        \addplot [
            color=red,
            mark=halfcircle
        ]
        coordinates{
        (0.01,0.0294)(0.02,0.0351)(0.03,0.0238)(0.04,0.0211)(0.05,0.0192)
        };
        \addlegendentry{$Gowalla$}
        \addplot [
            color=blue,
            mark=square
            ]
        coordinates{
        (0.01,0.0285)(0.02,0.0375)(0.03,0.0251)(0.04,0.0227)(0.05,0.0174)
        };
        \addlegendentry{$Foursquare$}
        \end{axis}
        \end{tikzpicture}}}
    \subfloat[][(c) $\lambda$]{%
		\label{fig:g-rec}%
		\resizebox{.25\textwidth}{!}{
		\begin{tikzpicture}
        \begin{axis}[
            axis lines = left,
            xlabel = $\lambda$,
            ylabel = {Recall@$20$},
            legend pos=south east,
            xmin=0, xmax=1,
            ymin=0, ymax=0.05,
            xtick={0,.1,.2,.3,.4,.5,.6,.7,.8,.9,1},
            xticklabels={0,.1,.2,.3,.4,.5,.6,.7,.8,.9,1},
            ytick={0,0.01,0.02,0.03,0.04,0.05}
        ]
        \addplot [
            color=red,
            mark=halfcircle
        ]
        coordinates{
        (0,0.0252)(0.1,0.0287)(0.2,0.0302)(0.3,0.0346)(0.4,0.0373)(0.5,0.0394)(0.6,0.0368)(0.7,0.0325)(0.8,0.0281)(0.9,0.0259)(1,0.0231)
        };
        \addlegendentry{$Gowalla$}
        \addplot [
            color=blue,
            mark=square
            ]
        coordinates{
        (0,0.0261)(0.1,0.0275)(0.2,0.0314)(0.3,0.0359)(0.4,0.0386)(0.5,0.0420)(0.6,0.0391)(0.7,0.0330)(0.8,0.0284)(0.9,0.0259)(1,0.0245)
        };
        \addlegendentry{$Foursquare$}
        \end{axis}
        \end{tikzpicture}}}
	\caption[]{Effect of different model parameters on the performance of \modelname}
	\label{fig:parameters}%
\end{figure}

\partitle{Effect of Model Parameters.}
Figure \ref{fig:parameters} shows the performance of \modelname for different values of $d$, $\alpha$ and $\lambda$. We report in Figure \ref{fig:d-pre} the effect of different values of $d$ on the performance of \modelname in terms of Precision@20 and Recall@20 metrics, respectively. It can be seen that the optimal value of $d$ for both datasets is 15. These results show that users tend to visit nearby POIs to their centers, which are formed in regions. Figure \ref{fig:a-pre}, on the other hand, shows the effect of different $\alpha$ values on the performance of \modelname. We can see that the optimal value is achieved at $\alpha = 0.02$ for both datasets. More importantly, as seen in Figure \ref{fig:g-pre}, the optimal value for $\lambda$ that shows the impact of different temporal states is $0.5$. This confirms our assumption and shows that users follow a spatio-temporal activity centered behavior. In fact, when we set this parameter to 1 (i.e., just working time) or 0 (i.e., just leisure time), the performance decreases.

\section{Discussion and Conclusion}
In this paper, we study the problem of POI recommendation. We have investigated in detail the characteristics of the user mobility behavior on a large-scale check-in dataset. Based on the extracted properties, we propose a novel spatio-temporal activity-centers model to jointly model the geographical and temporal influence of the user's check-in behavior. We then consider the user's temporal information and the user's preferences on POIs. Finally, we propose the \modelname model as a uniform framework for combining these three components to recommend POIs. Experimental results on the datasets show that our model significantly outperforms the state-of-the-art models. In our future work, we may consider more information such as the user's comments and social relations to improve the performance.

\partitle{Acknowledgment.}
This work was partially supported by a Swiss State Secretariat for Education, Research and Innovation (SERI) mobility grant between Switzerland and Iran.

\bibliographystyle{splncs04}
\bibliography{references.bib}

\end{document}